\documentclass[11pt,twoside]{article}
\usepackage{asp2014}

\aspSuppressVolSlug
\resetcounters

\hypersetup{urlcolor=blue, colorlinks=true, allcolors=blue}
\begin{document}

\newcommand{\astronomaly}{\texttt{Astronomaly}}

\title{A Hitchhiker's Guide to Anomaly Detection with \astronomaly{}}

\author{Michelle~Lochner$^{1,2}$ and Bruce.~A.~Bassett$^{3,4,5}$}
\affil{$^1$Department of Physics and Astronomy, University of the Western Cape, Bellville, Cape Town, 7535, South Africa; \email{mlochner@uwc.ac.za}}
\affil{$^2$South African Radio Astronomy Observatory (SARAO), 2 Fir Street, Observatory, Cape Town, 7925, South Africa}
\affil{$^3$African Institute for Mathematical Sciences, 6 Melrose Road, Muizenberg, 7945, South Africa}
\affil{$^4$Department of Maths and Applied Maths, University of Cape Town, Cape Town, South Africa}
\affil{$^5$South African Astronomical Observatory, Observatory, Cape Town, 7925, South Africa}

\paperauthor{Michelle~Lochner}{mlochner@uwc.ac.za}{0000-0003-2221-8281}{University of the Western Cape / South African Radio Astronomy Observatory}{Department of Physics and Astronomy}{Cape Town}{Western Cape}{7535}{South Africa}
\paperauthor{Bruce.~A.~Bassett}{bruce.a.bassett@gmail.com}{0000-0001-7700-1069}{African Institute for Mathematical Sciences}{Cosmology Group}{Cape Town}{Western Cape}{7945}{South Africa}



  
\begin{abstract}

The next generation of telescopes such as the SKA and the Rubin Observatory will produce enormous data sets, requiring automated anomaly detection to enable scientific discovery. Here, we present an overview and friendly user guide to the \astronomaly{} framework for active anomaly detection in astronomical data. \astronomaly{} uses active learning to combine the raw processing power of machine learning with the intuition and experience of a human user, enabling personalised recommendations of interesting anomalies. It makes use of a Python backend to perform data processing, feature extraction and machine learning to detect anomalous objects; and a JavaScript frontend to allow interaction with the data, labelling of interesting anomalous and active learning. \astronomaly{} is designed to be modular, extendable and run on almost any type of astronomical data. In this paper, we detail the structure of the \astronomaly{} code and provide guidelines for basic usage.
  
\end{abstract}

\section{Introduction}

Modern and future telescopes including MeerKAT\footnote{\url{https://www.sarao.ac.za/science/meerkat/about-meerkat/}}, the Square Kilometre Array\footnote{\url{https://www.skatelescope.org/}} and the Vera C. Rubin Observatory\footnote{\url{https://www.lsst.org/}} produce enormous quantities of data, making discovery of new and rare astrophysical phenomena impossible by manual data inspection. Machine learning has excellent potential for anomaly detection, but machine learning algorithms cannot distinguish between ``interesting'' anomalies and those that should be ignored. Furthermore, unlike classification problems where class definitions are generally agreed upon between scientists, different people will find different anomalies interesting. For example: an artefact may not be of interest to an astronomer looking for strong lenses but may be interesting to a pipeline scientist in charge of data quality.
To address this challenge, we developed \astronomaly{} \citep{2021A&C....3600481L} which is a general purpose framework for active anomaly detection in astronomical data. \astronomaly{} is publicly available on github\footnote{\url{https://github.com/MichelleLochner/astronomaly}}. We leave details of the framework and analysis of its performance to \citet{2021A&C....3600481L} and focus here instead on the implementation.

\section{Overview of \astronomaly{}}
\astronomaly{} is designed to be modular and flexible. It can either be used as a pure Python library for anomaly detection, or it can be used with the frontend to aid in active anomaly detection. It operates on the principle of ``code-as-config'' which allows maximum flexibility to operate with many data types, with the downside of being more complex to use.

\autoref{flowchart} shows an outline of \astronomaly{}, indicating the different Python and JavaScript components. It can be seen that the Python backend follows a standard machine learning pipeline structure. First data is read into a standard format and pre-processed as necessary. The next step is feature extraction, where complex data is simplified and summarised down to a smaller set of numbers. This is a crucial step and the choice of feature extractor will largely dictate the types of anomalies the algorithm will be sensitive to. If necessary, further dimensionality reduction can then be applied followed by any necessary post-processing. The features are then finally ready for a machine learning algorithm to be applied to detect anomalies. \astronomaly{} also allows the user to visualise the high dimensional feature space in two dimensions using embedding tools.

After this pipeline has been run, the JavaScript frontend allows the user to view the data from most to least anomalous, interact with it, assign scores based on how interesting each object is, and applying an active learning algorithm to improve the anomaly detection.

\articlefigure{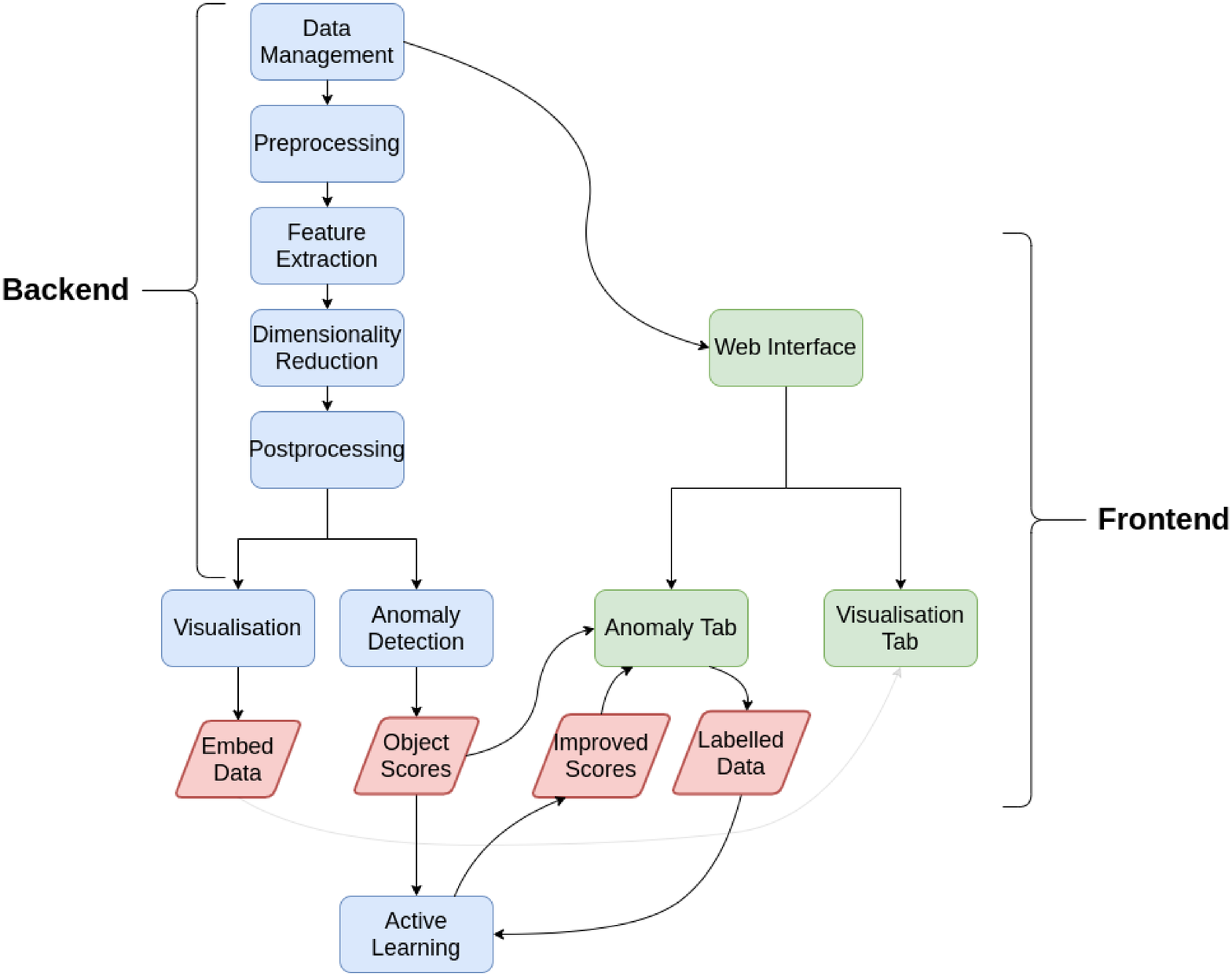}{flowchart}{Flowchart of \astronomaly{} software design. Rectangles indicate code components while the slanted parallelograms are outputs and inputs. The backend (blue) is built in Python while the frontend (green) is built in JavaScript, with a flask-based API connecting them \citep{2021A&C....3600481L}.}

\articlefigure{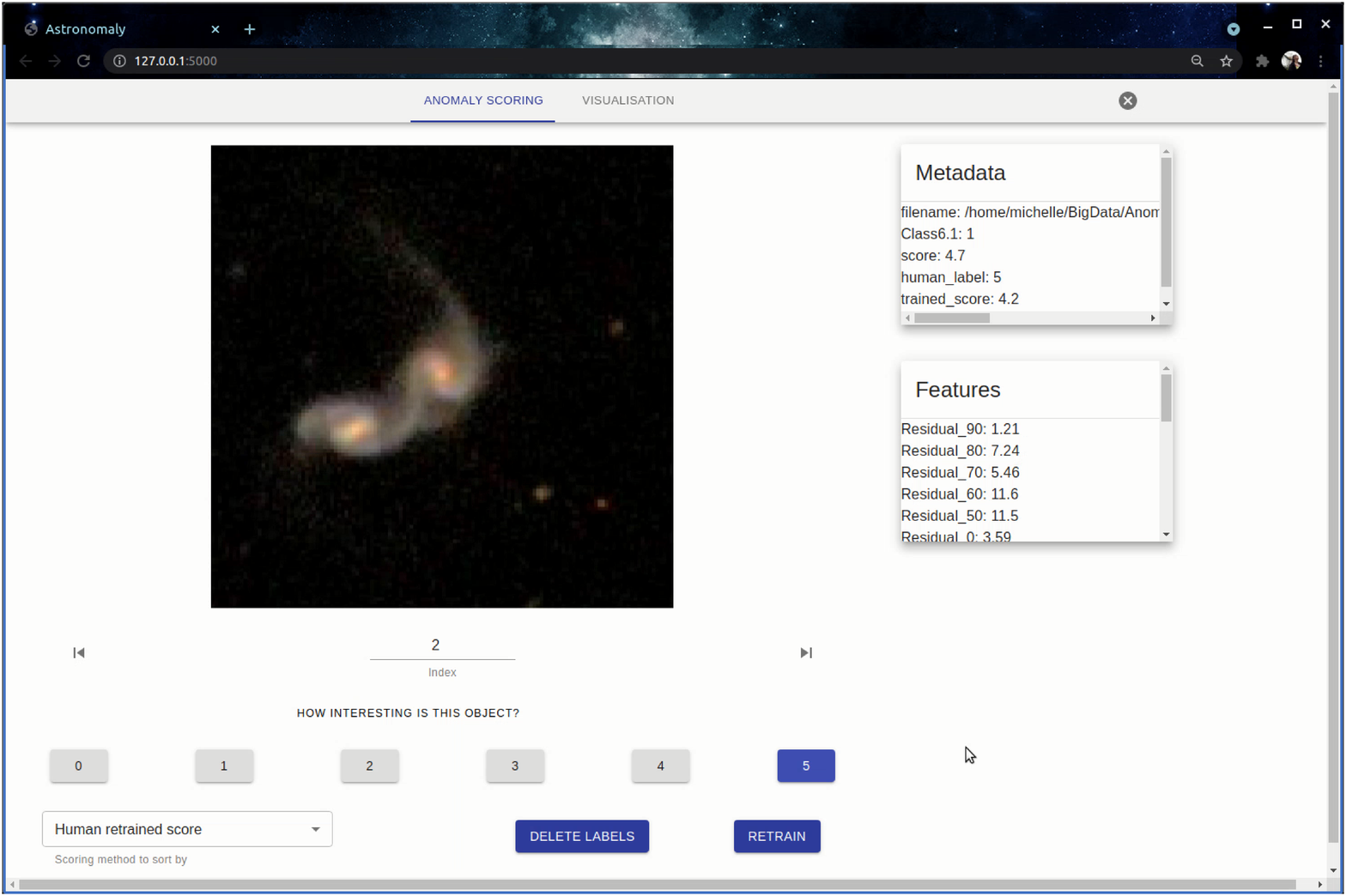}{screenshot}{Screenshot of \astronomaly{} frontend interface displaying an example galaxy. The numbered buttons allow the user to grade how interesting each anomaly is, the drop-down box allows changing how the data are ordered and the retrain button will call the active learning algorithm from the backend to improve the anomaly detection. Information about each object is displayed on the right. The visualisation tab changes the interface to an interactive 2d visualisation of feature space.}

\section{Example \astronomaly{} script}
\label{sec:script}
Several example scripts are included in the \astronomaly{} package. Here we step through the example script \verb!galaxy_zoo_example.py!, currently found in the \verb!scripts! folder of the github repository. This section may be best understood by opening this script and following it along with the text.

A simple Python script is how a user interacts with \astronomaly{}. Anything can be done in this script, allowing a user to (for example) use external tools to extract features or manipulate data. The only stipulation is that the script must provide a \verb!run_pipeline()! function which must return a dictionary containing (at minimum) the following keys:
\begin{itemize}
 \item \verb!'dataset'! - an object of type Dataset (further described in \autoref{sec:data})
 \item \verb!'features'! - a pandas \citep{mckinney-proc-scipy-2010, reback2020pandas} dataframe containing the final set of features, with the object identifier as the index and the features as columns (the column names will be assumed to be the feature names and used in the frontend)
 \item \verb!'anomaly_scores'! - a pandas dataframe containing the machine learning \\anomaly scores, with the object identifier as the index and the anomaly score as a column called \verb!'score'!
 \item \verb!'visualisation'! - a pandas dataframe containing the coordinates (i.e. two columns) for visualisation of the feature space, with the object identifier as the index
 \item \verb!'active_learning'! - an object that can run active learning on calling a function called \verb!'run()'! (see further details in \autoref{sec:active})
\end{itemize}

\subsection{Settings}
The first part of the \verb!galaxy_zoo_example.py! script simply sets up the file paths for the data and the output. \astronomaly{} by default stores the output of each stage of the pipeline as parquet\footnote{\url{https://parquet.apache.org/}} files. It will attempt to detect if a stage of the pipeline has already been run on the same data and, if it has, will load the stored output instead. This makes it easy to switch parts of the pipeline, such as experimenting with different machine learning algorithms, without needing to rerun slower parts such as feature extraction. This behaviour can be disabled at any stage of the pipeline by providing the keyword \verb!force_rerun=True!. Deleting the automatically generated log file, \verb!astronomaly.log! will also force a rerun of everything, since the log is what is used to detect any changes to pipeline stages.

\subsection{Reading in the data}
\label{sec:data}
In the \verb!run_pipeline()! function, usually the first step will be to load up the data. There are a variety of options in the \verb!data_management! package, allowing reading in of images (fits files and normal image file types), time series data and raw data such as spectra. The \verb!dev! branch of the github repository will generally contain the latest data readers available and it is straightforward to implement new ones (see \autoref{sec:extensions} and the online documentation for guidelines).

The line of code below reads in a set of image thumbnails into an object of type \verb!Dataset!, which the rest of \astronomaly{} is able to work with. The key functionality of objects of this type are access to an \verb!index! variable, which simply contains the list of object identifiers in the dataset, and two functions \verb!get_sample()! and \verb!get_display_data()! which both take an index and return the data for that object. In the latter case, the function must return a version of the data ready to be displayed by the JavaScript frontend. The \verb!Dataset! object must also contain a \verb!metadata! attribute which is a pandas dataframe with additional data about each object.

\begin{verbatim}
 image_dataset = image_reader.ImageThumbnailsDataset(
        directory=image_dir, output_dir=output_dir,
        transform_function=image_transform_function,
        display_transform_function=display_transform_function
    )
\end{verbatim}

Notice two important keywords, \verb!transform_function! and \\ \verb!display_transform_function!. These are functions or lists of functions that transform the data as it is read in. This is often a critical step in anomaly detection, allowing transforms like sigma clipping and non-linear flux transforms to be applied, highlighting critical features of the data to the algorithm. These functions will likely have a large impact on the success of your pipeline. In this script, they are defined earlier on and make use of a library of functions provided in the \verb!preprocessing! package. To avoid large up front computation, these transforms are called only when \verb!get_sample()! is called. A separate set of transform functions can be used for the visual display, to allow a user to, for example, see the original image instead of a sigma clipped image.

\subsection{Feature extraction}
After the data has been read in and pre-processed, the next step is to extract features. In this script, we use the same simple feature extractor used in \citet{2021A&C....3600481L}, which fits a series of ellipses to isophotes and uses the parameters as features. The method, which relies on OpenCV \citep{itseez2015opencv}, is very sensitive to how the data are pre-processed.

\begin{verbatim}
pipeline_ellipse = shape_features.EllipseFitFeatures(
    percentiles=[90, 80, 70, 60, 50, 0],
    output_dir=output_dir, channel=0, force_rerun=False,
    central_contour=False)

features = pipeline_ellipse.run_on_dataset(image_dataset)
\end{verbatim}

An object of type \verb!PipelineStage! is created first. While this is not essential (a user could for instance read in a set of features extracted elsewhere into a pandas dataframe), inheriting from the \astronomaly{} base class allows automated logging, rerun detection and in the future, parallelisation. The parameters for any feature extractor are passed as keyword arguments to the intialisation of the object. \verb!run_on_dataset()!, the function that actually performs the feature extraction, should only take a single argument, an object of type \verb!Dataset!. There are a variety of feature extractors to choose from in the \verb!feature_extraction! package.

\subsection{Feature post-processing}
In the \verb!galaxy_zoo_example.py! script, we then choose to scale the features to be centred around zero with unit variance. This is standard practice in machine learning and is a critical step for many (but not all) algorithms. The process for post-processing features is almost identical as that for extracting them, except you pass the function a pandas dataframe as input.

\begin{verbatim}
pipeline_scaler = scaling.FeatureScaler(force_rerun=False,
                                        output_dir=output_dir)
features = pipeline_scaler.run(features)
\end{verbatim}

\subsection{Machine learning}
We then run a machine learning algorithm for anomaly detection. Again, an external algorithm can be used as long as a correctly formatted pandas dataframe is returned from the \verb!run_pipeline()! function. In this example script, we make use of the popular isolation forest \citep{liu2008} algorithm to perform our initial anomaly detection.

\begin{verbatim}
pipeline_iforest = isolation_forest.IforestAlgorithm(
    force_rerun=False, output_dir=output_dir)
anomalies = pipeline_iforest.run(features)
\end{verbatim}
One can see that the procedure is the same as always: a \verb!PipelineStage! object (in this case \verb!IforestAlgorithm!) is created and then executed on the dataframe using the \verb!run()! function.

\astronomaly{} assumes the anomaly scores are of the type ``larger is more anomalous'', which is not the case for the output of isolation forest. Thus we make use of a simple converter to put the scores on a scale of 0 to 5, the same as the user interface.

\begin{verbatim}
pipeline_score_converter = human_loop_learning.ScoreConverter(
    force_rerun=False, output_dir=output_dir)
anomalies = pipeline_score_converter.run(anomalies)
\end{verbatim}

\subsection{Active learning}
\label{sec:active}
In the \verb!galaxy_zoo_example! script, there is a small set of code which checks if labelling has already been done. This does not need to be changed generally.

In order to perform active learning, something slightly different to the previous steps must be done. An object, of type \verb!PipelineStage!, must be created which is capable of executing an active learning algorithm and returning a dataframe with the new scores. \astronomaly{} currently provides one active learning algorithm, as described in \citet{2021A&C....3600481L}, which is created as follows:

\begin{verbatim}
pipeline_active_learning = human_loop_learning.NeighbourScore(
        alpha=1, output_dir=output_dir)
\end{verbatim}

In \astronomaly{}, we assume that the active learning algorithm can take in a set of features, the original anomaly score (from the machine learning algorithm) and a set of human applied labels, and then return a corresponding dataframe with a new retrained score. The frontend needs access to the actual object, in this case the \verb!NeighbourScore! object, to be able to run the active learning on demand. This differs from the other stages of the pipeline (such as feature extraction or machine learning) which are static and do not change as the user interacts with \astronomaly{}.

\subsection{Visualisation}
Finally, we choose to enable visualisation of our feature space using t-SNE \citep[][although any other similar algorithm can be used]{vandermaaten2008}, which is called in much the same way as the previous steps.

\begin{verbatim}
pipeline_tsne = tsne_plot.TSNE_Plot(
        force_rerun=False,
        output_dir=output_dir,
        perplexity=100)
t_plot = pipeline_tsne.run(features)
\end{verbatim}

\subsection{Return requirements}
It is very important that a return statement is included in the \verb!run_pipeline()! function with all the required keys in the dictionary.
\begin{verbatim}
return {'dataset': image_dataset,
        'features': features,
        'anomaly_scores': anomalies,
        'visualisation': t_plot,
        'active_learning': pipeline_active_learning}
\end{verbatim}

\subsection{Running \astronomaly{}}
The script is then passed directly to the \verb!run_server! script as follows:

\begin{verbatim}
python frontend/run_server.py scripts/galaxy_zoo_example.py
\end{verbatim}

The pipeline will execute and display some text when it is finished:
\begin{verbatim}
 ##### Astronomaly server now running #####
Open this link in your browser: http://127.0.0.1:5000/
\end{verbatim}
Opening that link with a browser will then display the \astronomaly{} frontend, similar to \autoref{screenshot}.

\section{The Frontend}
After calling \verb!run_server()! and inputting a script, \astronomaly{} will launch the frontend. This is built using JavaScript, primarily making use of the React\footnote{\url{https://reactjs.org/}} framework, allowing easy extendibility. The frontend display is split into two tabs: the anomaly tab and the visualisation tab.

\subsection{Anomaly tab}
As can be seen in \autoref{screenshot}, the anomaly tab allows the user to interact with the data and score objects based on how interesting they are. The data can be ordered several different ways: randomly (to get a sense for the average objects in the dataset), according to the machine learning score and (after active learning is applied) according to the retrained score. The ``retrain'' button allows the active learning algorithm to be called after labelling some data and the ``delete labels'' button can be used, with caution, to reset the labels that would otherwise be saved between sessions.

\subsection{Visualisation tab}
\autoref{screenshot_tsne} shows the visualisation tab of the \astronomaly{} interface where a large fraction of the data set can be visualised, in this case using t-SNE (although any method of reducing the data down to two dimensions can be used and is changed in the python script). This tool is helpful to learn patterns in the data and evaluate how effective the features are at not only detecting anomalies, but also grouping similar anomalies together. The type of plot shown on the right will change depending on the data type used (see \autoref{sec:api} for a discussion on how the plots are generated).
\articlefigure{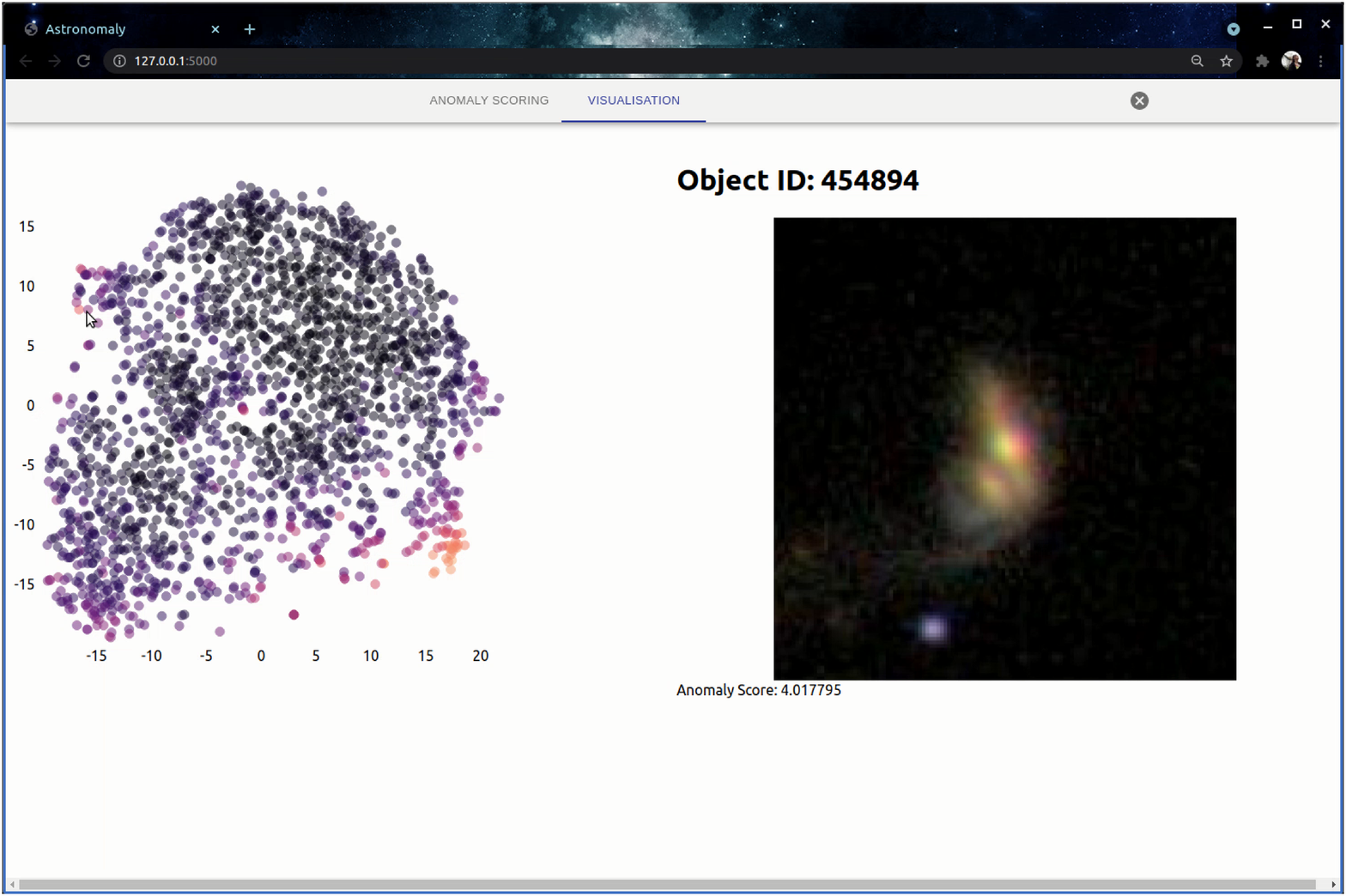}{screenshot_tsne}{Screenshot of the visualisation tab of the \astronomaly{} frontend interface. The features are visualised using some form of dimensionality reduction or manifold embedding, in this case t-SNE. Each point on the plot in the left represents a different object in the dataset. The points are coloured according to anomaly score (lighter is more anomalous). Clicking on a point on the plot will display the data for that object.}

\subsection{REST API}
\label{sec:api}
To connect the Python backend with the JavaScript frontend, we built a REST API using Flask\footnote{\url{https://flask.palletsprojects.com/en/2.0.x/}}. A REST (Representational state transfer) API (application programming interface) is a simple set of rules of communication between two programs, making use of requests and responses.

The key components of \astronomaly{}'s REST API are found in the \verb!frontend! package. The \verb!interface! module implements a \verb!Controller! class which facilitates calls to the backend. It takes the Python script (such as the one discussed in \autoref{sec:script}), executes the pipeline and then has several other functions designed to be called from the frontend.

The \verb!run_server! module implements the key REST API functionality. It uses Flask to generate requests to the frontend and handle responses. For example, when a user changes the object they are viewing, the frontend generates a request, that contains the index of the new object, which gets passed to \verb!get_image()! function in the \verb!run_server! module. That function posts a response, which is the actual image in bytes that can be displayed directly on the web page.

The type of data (images, time series or raw features) is stored as an attribute in the Python \verb!Dataset! object and read in when \astronomaly{} is called. The data type will dictate the kinds of plots the frontend will display. This flexibility is implemented in the \verb!PlotContainer.js! module which will use different components for different data types. This makes it easy to use the same code to work with a variety of astronomical data and to extend the frontend to include other types of plots.

\section{Extending \astronomaly{}}
\label{sec:extensions}
\astronomaly{} is designed to be modular and flexible, and hence easy to extend. As mentioned before, because \astronomaly{} relies on a Python script and not on predefined controls, one can insert almost any kind of external analysis. However, it is straightforward to incorporate a new element into the \astronomaly{} library.

\astronomaly{} makes extensive use of \emph{inheritance} to allow new classes to inherit a standard set of behaviours such as automated logging and automatically detecting if a pipeline stage has already been run. There are two types of base classes in \astronomaly{}: the \verb!Dataset! class, and the \verb!PipelineStage! class. If the user has a new dataset to be read into \astronomaly{} which cannot be read by an existing data reader in the \verb!data_management! package, a new class of type \verb!Dataset! must be created. If however, the user requires a new feature extractor, machine learning algorithm or essentially any other new step in the pipeline, a new class must be created which inherits from \verb!PipelineStage!.

Technical details of how to extend \astronomaly{} can be found in the online documentation: \url{https://astronomaly.readthedocs.io/en/latest/contributing.html}

\section{Conclusions}
We have discussed some of the implementation details of the \astronomaly{} package - a framework for personalised anomaly detection in astronomical data. With a modular Python backend, JavaScript frontend and REST API connecting the two, \astronomaly{} is flexible enough to be extended to include new data types, feature extractors, machine learning algorithms and frontend components.

\astronomaly{} is still being actively developed. We plan to implement easy functionality to search a variety of online databases by coordinates to facilitate quick follow-up of interesting sources. We will also include a clustering component for rapid labelling of data, similarity searches and better export of the anomalies found. Finally, development into more effective feature extractors and machine learning algorithms is ongoing, and contributions from the community are more than welcome.

\bibliography{I7-001}


\end{document}